\documentclass[lettersize,journal]{IEEEtran}
\usepackage{amsmath,amsfonts}
\usepackage{algorithmic}
\usepackage{array}
\usepackage[caption=false,font=normalsize,labelfont=sf,textfont=sf]{subfig}
\usepackage{textcomp}
\usepackage{stfloats}
\usepackage{url}
\usepackage{verbatim}
\usepackage{graphicx}
\usepackage{cleveref}
\usepackage{cite}
\def\BibTeX{{\rm B\kern-.05em{\sc i\kern-.025em b}\kern-.08em
    T\kern-.1667em\lower.7ex\hbox{E}\kern-.125emX}}
\usepackage{balance}
\begin{document}
\title{Solar Cells, Lambert W and the LogWright Functions}
% \author{
%   \IEEEauthorblockN{
%     Prabhat Lankireddy\IEEEauthorrefmark{1},
%     Sibibalan Jeevanandam\IEEEauthorrefmark{2},
%     Aditya Chaudhary\IEEEauthorrefmark{3},
%     P. C. Deshmukh\IEEEauthorrefmark{4},
%     Ken Roberts\IEEEauthorrefmark{5},
%     and
%     S. R. Valluri\IEEEauthorrefmark{5}\IEEEauthorrefmark{6}\\}
%   \IEEEauthorblockA{\IEEEmembership{\IEEEauthorrefmark{1}Centre for Machine Intelligence and Data Science, Indian Institute of Technology Bombay (IITB), Mumbai, India\\}}
%   \IEEEauthorblockA{\IEEEmembership{\IEEEauthorrefmark{2}Dept. of Mechanical Engineering, Purdue University, Indiana, U.S.A\\}}
%   \IEEEauthorblockA{\IEEEmembership{\IEEEauthorrefmark{3}Dept. of Mechanical Engineering, Indian Institute of Technology Tirupati (IITT), Tirupati, India\\}}
%   \IEEEauthorblockA{\IEEEmembership{\IEEEauthorrefmark{4}Dept. of Physics and CAMOST, Indian Institute of Technology Tirupati (IITT), Tirupati, India\\}}
%   \IEEEauthorblockA{\IEEEmembership{\IEEEauthorrefmark{5}Department of Physics and Astronomy, University of Western Ontario (UWO), London, ON, Canada\\}}
%   \IEEEauthorblockA{\IEEEmembership{\IEEEauthorrefmark{6}Department of Management, Economics, and Mathematics, Kings University College, UWO, London, ON, Canada\\}}
% }
\author{
    Prabhat Lankireddy,
    Sibibalan Jeevanandam,
    Aditya Chaudhary,
    P. C. Deshmukh,
    Ken Roberts,
    and S. R. Valluri
\thanks{\emph{Corresponding Author: S. R. Valluri (e-mail: valluri@uwo.ca)}}
\thanks{P. Lankireddy is with the Dept. of Electrical Engineering, Indian Institute of Technology Tirupati (IITT),
Tirupati, India, and the Center for Machine Intelligence and Data Science (C-MInDS), Indian Institute of Technology Bombay (IITB), Mumbai, India.}
\thanks{S. Jeevanandam is with the Dept. of Mechanical Engineering, Indian Institute of Technology Tirupati (IITT),
Tirupati, India, and the Dept. of Mechanical Engineering, Purdue University, Indiana, U.S.A.}
\thanks{A. Chaudhary is with the Dept. of Mechanical Engineering, Indian Institute of Technology Tirupati (IITT),
Tirupati, India.} %Not sure about Aditya's affiliation
\thanks{P. C. Deshmukh is with the Dept. of Physics and CAMOST, Indian Institute of Technology Tirupati (IITT), Tirupati, India and the Dept. of Physics, Dayananda Sagar University, Bengaluru, India.}
\thanks{K. Roberts is with the Department of Physics and Astronomy, University of Western Ontario (UWO), London,
ON, Canada.}
\thanks{S. R. Valluri is with the Department of Physics and Astronomy and the Department of Management, Economics and Mathematics, King's College, University of Western Ontario (UWO), London, ON, Canada.}
}

% \markboth{IEEE Journal of Photovoltaics,~Vol.~18, No.~9, September~2020}%
% {How to Use the IEEEtran \LaTeX \ Templates}
% \markboth{Preprint}%
\markboth{Preprint to ArXiv}%
{Prabhat Lankireddy, Sibibalan Jeevanandam, Aditya Chaudhary, P. C. Deshmukh, Ken Roberts, and S. R. Valluri, \MakeLowercase{\textit{(et al.)}: Solar Cells, Lambert W and the LogWright Functions}}

\maketitle

\begin{abstract}
Algorithms that calculate the current-voltage (I-V) characteristics of a solar cell play an important role in processes that aim to improve the efficiency of a solar cell. I-V characteristics can be obtained from different models used to represent the solar cell, and the single diode model is a simple yet accurate model for common field implementations. However, the I-V characteristics are obtained by solving implicit equations, which involve repeated iterations and inherent errors associated with numerical methods used. Some methods use the Lambert W function to get an exact explicit formula, but often causes numerical overflow problems. The present work discusses an algorithm to calculate I-V characteristics using the LogWright function, a transformation of the Lambert W function, which addresses the problem of arithmetic overflow that occurs in the Lambert W implementation. An implementation of this algorithm is presented and compared against other algorithms in the literature. It is observed that in addition to addressing the numerical overflow problem, the algorithm based on the LogWright function offers speed benefits while retaining high precision.
\end{abstract}

\begin{IEEEkeywords} %TODO: Need to verify these keywords
  Solar Cell, Photovoltaic Cell, Single Diode Model, Lambert W function, LogWright function, I-V characteristics
\end{IEEEkeywords}

\section{Introduction}
\IEEEPARstart{S}{olar} cells have been extensively used for decades to convert light energy into electricity through the photovoltaic effect. They are one of the major sustainable options for generating electricity, owing to its benefits like long equipment lifespan and absence of harmful emissions. Solar cells find their applications in electric cars, water-heating, lighting systems, telecommunication towers and so on. \\
A solar panel comprises many solar cells arranged in series and parallel configurations. The output of the individual cells may vary due to shading in the panel, or due to failure or poor performance of some cells due to aging. A defect in a string (series) of cells can cause loss of current through the whole string. This can be due to a shadow, not just a semi-permanent characteristic of the cell. For instance, a building chimney can shade a part of a roof panel, with the shadow moving slowly across the panel during the day. Shading can change even faster in case of a car with a panel on its roof being driven along a tree-lined avenue. Dust in the air, swirling sand deposited on a roof panel, can cause even more rapid changes in shading. Some parts of the panel receive full sunlight and keep pumping out energy which must go somewhere. When not provided with a path to flow, that energy gets dissipated as heat, possibly damaging the cells. This brings the need to perform real-time load balancing in order to minimize losses. To perform such calculations, it is essential to develop mathematical models and simulate the behavior of solar cells using the same. \\

Solar cells are modeled in different ways based on the diode configuration. The single diode model is extensively used to represent solar cells, owing to its simplicity and accuracy in many cases. The single diode model is described by an implicit equation which relates the current I and the voltage V in terms of the cell parameters as described in literature\cite{nelson2003solarcells}. Active load balancing aims at establishing the optimal operation (at the maximum power point, when the product of current and voltage is maximized). Load balancing requires calculations, and in some cases, there is a need to solve an implicit equation by repeated iteration. Note that mathematical models used to characterize the voltage-current relationship of solar cells are based on approximations. Solar cells are not really single diode circuits with a series and a bypass resistor. Therefore, a practical implementation should also involve a history of past behaviour of the panel as experienced in the field. Ideally, each field installation would have its own computer which runs the model, keeps history, and communicates with the load equipment. These field installations consist of special-purpose hardware with small compute power, and using explicit formulas to calculate the I-V characteristics of the Solar Cell is much more efficient that using implicit formulas. Moreover, large errors in the output due to approximations while solving an implicit equation\cite{charles1981bluegrey} are also avoided.\\

Jain and Kapoor\cite{JAIN2004269} gave an exact explicit formula for V as a function of I using the Lambert W function. Since then, many publications have proposed the use of the Lambert W function for tasks such as extraction of I-V characteristics, parameter estimation and maximum power point tracking. This brought about the need to develop fast and efficient algorithms to evaluate the Lambert W function. Batzelis et al.~\cite{batzelisComputationLambertFunction2020} compare a variety of explicit non-iterative methods that use approximations to quickly evaluate the Lambert W function. Toledo et al.~\cite{toledo2022quick} propose using an approximation of the Lambert W as a seed for the iterative process, so that one can choose to use the approximation as is and trade-off speed with precision if needed. Nguyen et al.~\cite{nguyenSolarPVModeling2022} propose a convex programming approach to accurately evaluate the Lambert W function. \\

However, existing literature reports arithmetic overflow exceptions when using the Lambert W function to evaluate I-V characteristics. For instance, at higher shunt resistance (used for enhancing the efficiency of solar cells), for a current $I=0$ (open-circuit), the argument of the W-function can be $4.59 \times 10^{1141}$. However, the maximum representable magnitude in IEEE-754-2008 compliant double precision is only about $10^{323}$\cite{1984}! This makes it difficult for this formula to be used in hardware implementations of load balancing algorithms. To get around this issue, Batzelis et al. \cite{batzelisComputationLambertFunction2020} propose a way to evaluate the Lambert W function by expressing the function argument as $x=a e^{b}$ and using the parameters $a$ and $b$ instead of using $x$ to evaluate the function. While such an implementation solves the overflow issue, it would not help when the user wishes to calculate the derivative of the I-V characteristic.\\

% Addition needed: Need to discuss other relevant literature that propose similar methods. Need to bring up Batzelis, Toledo, and the cone programming approach.
Hence, it is desirable to replace the equation of Jain and Kapoor with an equation that: (i) is exact, analytic, explicit; (ii) is robust when used in calculations; (iii) does not pose a risk of arithmetic overflow; (iv) is desirable for lab studies, industrial applications, and field installations (load balancing); (v) is suitable for programming in Fortran or C or for microcontrollers; and (vi) reduces possible cancellations which cause loss of significant digits.
% Can discuss methods discussed in other publications here.
% Toledo et al. and Batzelis et al. get around the numerical overflow problem by expressing the argument $x=a e^b$ and using the parameters $a$ and $b$ instead of using $x$ whenever $x$ is very large.
A method proposed by Roberts and Valluri\cite{valluri2016solar} that uses the LogWright function, also called the \textit{g-function}, helps address the overflow problem caused by the Lambert W function and satisfies all the requirements posed above. The LogWright function has previously been used for prediction of conversion efficiency and real-time maximum power in photovoltaic modules\cite{ZAIMI2020113071}.\\

In this paper, an approach to calculate the I-V characteristics of a solar cell using the LogWright function is discussed and compared with other methods proposed in the literature that use the Lambert W function.
In \cref{sec:single-diode-models}, the single diode model is described and both the implicit and explicit equations that relate the output current and voltage are presented. In \cref{sec:logwright-sdm}, the LogWright function is introduced and the SDM equations in terms of the LogWright function are obtained. An algorithm to evaluate the LogWright function, which was proposed by Roberts~\cite{roberts2015robust}, is discussed in \cref{subsec:logwright-algorithm}. Two approaches to compute the I-V characteristics using LogWright function are discussed in \cref{sec:comparing-algorithms}, and the performance of their implementations is compared with other algorithms proposed in the literature. Lastly, \cref{sec:conclusion} concludes the comparative study and lists the scope for future work.

\section{Single Diode Model of Solar Cell}
\label{sec:single-diode-models}
\begin{figure}[!t]
    \centering
    \includegraphics[width=3.2in]{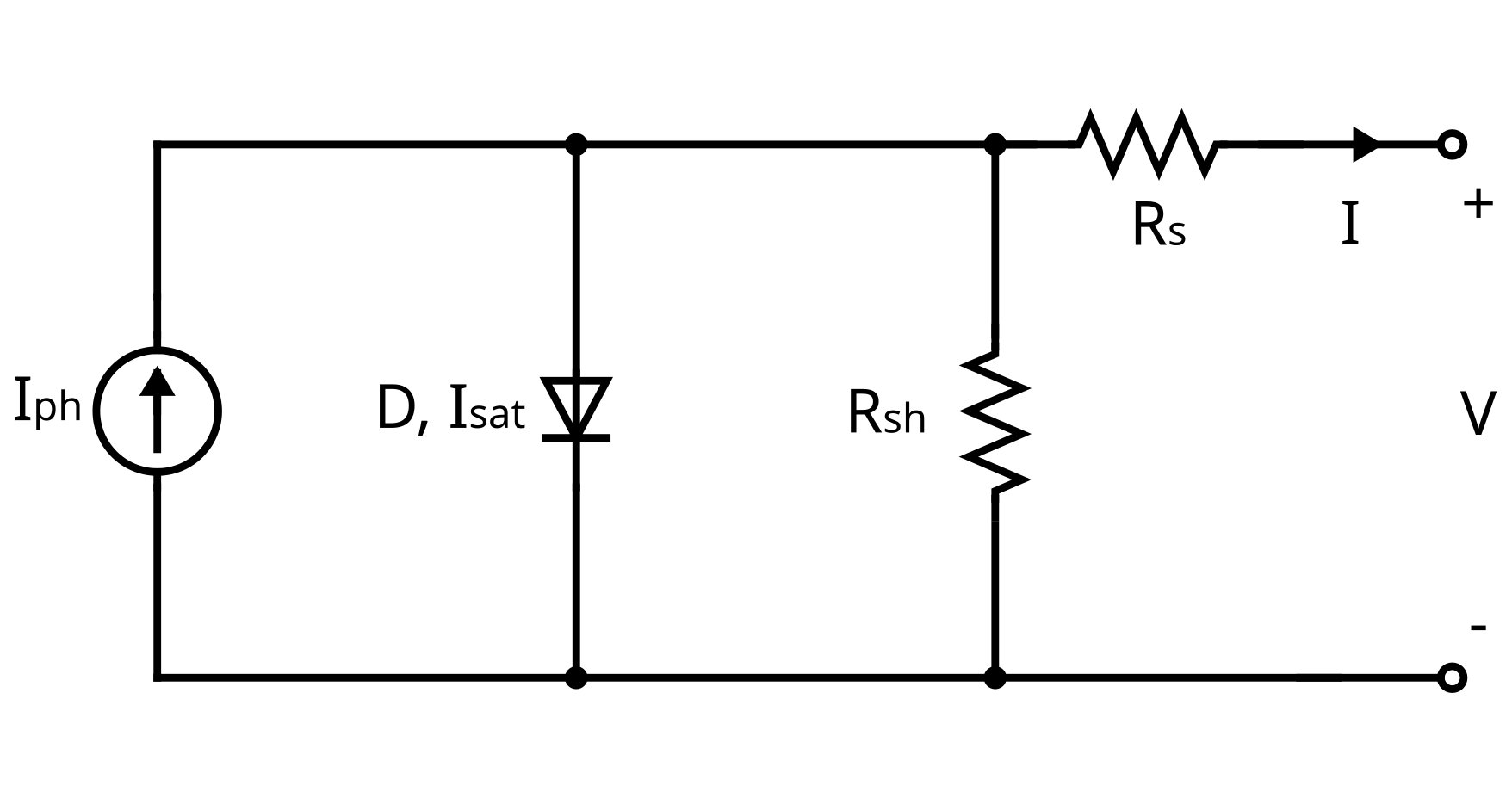}
    \caption{Circuit representing the Single Diode Model of a Solar Cell}
    \label{fig:solar-cell-sdm}
\end{figure}

\noindent The Single Diode Model (SDM) represents a solar cell using a constant current source, a diode, a series and a bypass resistor, and it approximates an inorganic solar cell quite well. The schematic of the SDM is shown in \cref{fig:solar-cell-sdm}. Modeling a solar panel using the SDM involves adjusting the model parameters, which include resistance and current values, such that the I-V characteristics of the model match that of the solar panel.
%\todo{I suggest replacing "Single Diode Model" with "SDM" everywhere after the first sentence}

A solar panel with a string of $N_s$ cells is modelled using the SDM, and the case of a single cell is handled as $N_s=1$.  In either situation, the output current of the solar cell $I$ is given by \cref{eq:i-implicit}\cite{nelson2003solarcells}.
\begin{equation}
\label{eq:i-implicit}
  I = I_{ph} - I_{sat}\left(\exp{\left(\frac{V+IR_s}{a}\right)} - 1\right) - \frac{V+IR_s}{R_{sh}}
\end{equation}
where $V$ is the output voltage of the solar panel, $I_{ph}$ is the photocurrent of the panel, $I_{sat}$ is the saturation current of the diode in the SDM and $a$ is a constant defined as $a=\eta N_s V_{Th}$. Here, $\eta$ is the ideality factor of the diode, and $V_{Th}=\frac{kT}{q}$ is the thermal voltage ($k$ is the Boltzmann constant, $T$ is the temperature and $q$ is the charge of an electron). One can obtain an explicit expression for I using the Lambert W function~\cite{JAIN2004269}, which can be used to calculate I if all other parameters are known.
\begin{multline}
\label{eq:i-explicit}
  I = \frac{1}{R_{sh}+R_s}(R_{sh}(I_{ph}+I_{sat})-V) \\ - \frac{a}{R_s} W_0 \left( \frac{I_{sat}R_{sh}R_{s}}{a(R_{sh}+R_s)} \exp\left(\frac{R_{sh}(R_s(I_{ph}+I_{sat})+V)}{a(R_{sh}+R_s)}\right) \right)
\end{multline}

Similarly, an explicit expression can be obtained for the output current $V$ as well.
\begin{multline}
\label{eq:v-explicit}
  V = R_{sh}(I_{ph}+I_{sat}) - (R_{sh}+R_s)I \\ -aW_0\left( \frac{I_{sat}R_{sh}}{a} \exp\left( \frac{R_{sh}}{a}(I_{ph}+I_{sat}-I) \right) \right)
\end{multline}

Either of \cref{eq:i-explicit} or \cref{eq:v-explicit} can be used to obtain the I-V characteristics of the solar panel, by manually varying $V$ or $I$ respectively. However, such computations are often impractical as numerical overflows tend to occur during calculations. For example, the argument of the Lambert W function in \cref{eq:v-explicit} can reach the order of $10^{1100}$, which cannot be represented by a IEEE-754-2008 standard hardware floating point in double precision (which has a limit around $10^{323}$). Such overflows occur due to the presence of exponentials in the argument of the Lambert W function. Such an overflow problem can be addressed using the LogWright function.

\section{The LogWright function}
\label{sec:logwright-sdm}
\noindent The LogWright function, also called the g-function\cite{roberts2015calculating,roberts2015robust}, is the logarithm of the Wright $\omega$ function~\cite{corlessWrightOmegaFunction2002}. It addresses the overflow problem by mathematically modifying the Lambert W function such that intermediate arguments with large numerical magnitudes are avoided. The following shows the relation between the LogWright function $g(x)$ and the Lambert W function.
\begin{equation}
    \label{eq:g-function}
    g(x) = \log{(W(e^x))} = x - W(e^x).
\end{equation}

\begin{figure}[!t]
    \centering
    \includegraphics[width=3.2in]{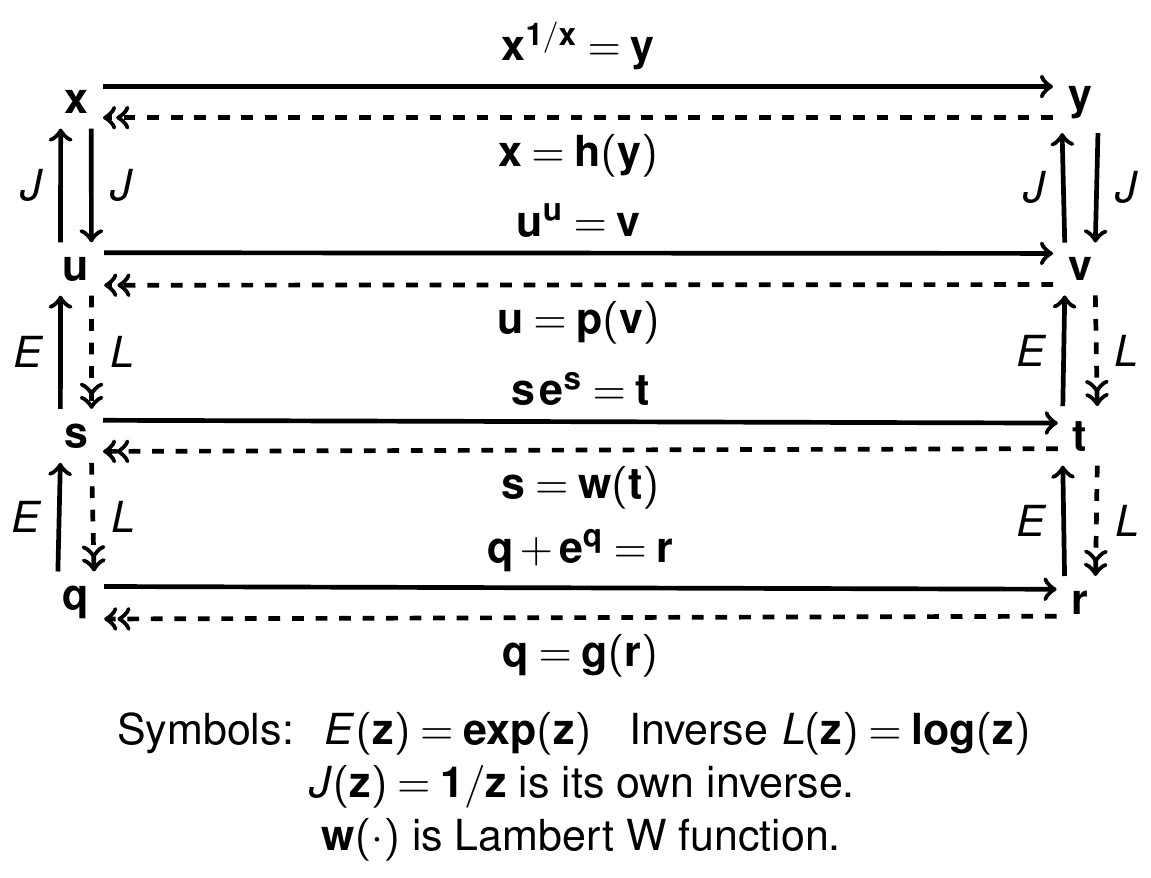}
    \caption{The Lambert W Function Ladder\cite{valluri2016solar}\cite{valluriSolarCellsCAP2022}}
    \label{fig:lambert-w-ladder}
\end{figure}

Note that $\log$ denotes the natural logarithm (to base $e$). The LogWright function can be obtained from the Lambert W function by taking the logarithm of the function parameters. However, a more efficient way to compute the LogWright function is discussed in \cref{subsec:logwright-algorithm}. This logarithmic transformation prevents extremely large numbers from occurring during intermediate calculations. The Lambert W ladder\cite{valluri2016solar}\cite{valluriSolarCellsCAP2022} shown in \cref{fig:lambert-w-ladder} gives a better understanding of how the LogWright function is obtained from the Lambert W function. In this diagram, the dashed lines with double arrowheads represent multi-valued functions.

Eq.~\ref{eq:i-explicit} can be rewritten in terms of the LogWright function as
\begin{equation}
  \label{eq:i-gfunction}
  I = \frac{a}{R_{s}} \left( g(u(V)) - \log{\left( \frac{I_{sat}R_{s}}{a(1+\frac{R_s}{R_{sh}})} \right)} \right) - \frac{V}{R_s}
\end{equation}
% $a=\eta N_s V_{Th}$, $\eta$ is the ideality factor, $N_s$ is the number of cells in series, $V_{Th}=\frac{kT}{q}$ is the thermal voltage, $R_{sh}$ is the shunt resistance\todo{weren't these (except for $u(V)$) already specified in the previous section?}, and

where $u(V)$ is
\begin{equation}
  \label{eq:i-gfunction-aux}
  u(V) = \log{\left( \frac{I_{sat}R_{s}R_{sh}}{a(R_s + R_{sh})} \right)} + \frac{R_s R_{sh}(I_{PV} + I_{sat}) + V R_{sh}}{a(R_{sh}+R_s)}.
\end{equation}

Similarly \cref{eq:v-explicit} can be rewritten as
\begin{equation}
  \label{eq:v-gfunction}
  V = a \left( g(v(I)) - \log{\left(\frac{I_{sat}R_{sh}}{a}\right)} \right) - IR_s
\end{equation}

where $v(I)$ is
\begin{equation}
  \label{eq:v-gfunction-aux}
  v(I) = \log{\left( \frac{I_{sat}R_{sh}}{a} \right) + \frac{(I_{PV}+I_{sat}-I)R_{sh}}{a}}.
\end{equation}

Note that certain terms in \cref{eq:i-gfunction,eq:i-gfunction-aux,eq:v-gfunction,eq:v-gfunction-aux} have different signs from the equations derived by Roberts and Valluri \cite{roberts2015calculating}, owing to the reversal in the direction of output current convention (as seen in \cref{fig:solar-cell-sdm}).

% \section{Algorithm to calculate I-V characteristics using the LogWright function}
\subsection{Algorithm to evaluate the LogWright function}
\label{subsec:logwright-algorithm}

\noindent The LogWright function is evaluated using the following 3-step process, as proposed by Roberts and Valluri~\cite{roberts2015robust}:
\begin{enumerate}
    \item Make an initial (crude) estimate of $y_0$ in the following manner:
    \begin{itemize}
        \item If $x \leq -e$, take $y_0=x$.
        \item If $x \geq e$, take $y_0=\log{x}$.
        \item If $-e \leq x \leq e$, take $y_0 = -e + \frac{1+e}{2e}(x+e)$ (linear interpolation between points (-e,-e) and (e,1)).
    \end{itemize}

    \item Refine the estimate $y_0$ by calculating
    \begin{equation*}
        y_1 = y_0 - \frac{2(y_0+e^{y_0}-x)(1+e^{y_0})}{2(1+e^{y_0})^2-(y_0+e^{y_0}-x)e^{y_0}}
    \end{equation*}
    This iteration formula is referred to as Halley's method and has cubic convergence (Note that the Newton Raphson method was used in \cite{toledo2022quick}, which has quadratic convergence).
    \item Perform additional iterations to get a better approximation
    \begin{equation*}
        y_n = y_{n-1} - \frac{2(y_{n-1}+e^{y_{n-1}}-x)(1+e^{y_{n-1}})}{2(1+e^{y_{n-1}})^2-(y_{n-1}+e^{y_{n-1}}-x)e^{y_{n-1}}}
    \end{equation*}

\end{enumerate}
In the interest of computational efficiency, it is desirable to calculate $e^{y_0}$ in step 2 and $e^{y_{n-1}}$ in step 3 only once and reuse the calculated values. This prevents repeated calls to the exponential function and speeds up the algorithm. Also, one can avoid unnecessary iterations with careful design in order to improve efficiency. In most practical application situations, only one iteration of step 3 is needed as $y_2$ is accurate to about 8 significant digits. This depends upon the voltage levels present in the equipment.

% The author names and affiliations could be formatted in two ways:
% \begin{enumerate}[(1)]
% \item Group the authors per affiliation.
% \item Use footnotes to indicate the affiliations.
% \end{enumerate}
% See the front matter of this document for examples. You are recommended to conform your choice to the journal you are submitting to.

\begin{table}
    \centering
    \caption{Parameter values of different single diode models considered for the comparative study.}
    \label{tab:model-parameter-values}
    \begin{tabular}{cccc}
      \hline
      \textbf{Parameters (Units)} & \textbf{Set 1} & \textbf{Set 2} & \textbf{Set 3} \\
      \hline
      $I_{ph}~(A)$ & 15.88 & 1.032 & 3.654 \\
      $I_{sat}~(A)$ & $7.44 \times 10^{-10}$ & $2.513 \times 10^{-6}$ & $3.999 \times 10^{-21}$ \\
      $R_{s}~(\Omega)$ & 2.04 & 1.239 & 2.69 \\
      $R_{sh}~(\Omega)$ & 425.2 & 744.714 & 2329 \\
      $a~(V)$ & 14.67 & 1.3 & 0.516 \\
      $I_{sc}~(A)$ & 15.804 & 1.031 & 3.650 \\
      $V_{oc}~(V)$ & 348.1 & 16.775 & 24.893 \\
      \hline
      \hline
      \textbf{Parameters (Units)} & \textbf{Set 4} & \textbf{Set 5} & \textbf{Set 6} \\
      \hline
      $I_{ph}~(A)$ & 0.578 & 0.761 & 4.802 \\
      $I_{sat}~(A)$ & $1.34 \times 10^{-10}$ & $3.107 \times 10^{-7}$ & $4.016 \times 10^{-7}$ \\
      $R_{s}~(\Omega)$ & $0.0127$ & $0.037$ & $0.5906$ \\
      $R_{sh}~(\Omega)$ & 612 & 52.89 & 1167 \\
      $a~(V)$ & 0.0118 & 0.039 & 0.037 \\
      $I_{sc}~(A)$ & 0.578 & 0.760 & 1.006 \\
      $V_{oc}~(V)$ & 0.262 & 0.573 & 0.603 \\
      \hline
    \end{tabular}
\end{table}

\section{Experiments and Results}
\label{sec:comparing-algorithms}
\noindent One can obtain the I-V characteristics of a single diode model using the LogWright function by following one of the two approaches:
\begin{enumerate}
  \item \textbf{I-approach}: Solve for current by using \cref{eq:i-gfunction} and varying the voltage values.
  \item \textbf{V-approach}: Solve for voltage by using \cref{eq:v-gfunction} and varying current values.
\end{enumerate}
Both the above approaches require evaluating the LogWright function. These approaches, along with the algorithm to evaluate the LogWright function, are implemented in the Python programming language and compared with other methods used to evaluate I-V characteristics. We compare the LogWright-based approaches against the following methods.
\begin{itemize}
  \item The SciPy\cite{2020SciPy-NMeth} implementation of the Lambert W function, which is based on the algorithm proposed by \cite{corless1996lambert}.
  \item Approach proposed by Toledo et al.~\cite{toledo2022quick}. This algorithm chooses between two different variants of the Lambert W function depending upon the value of the argument $x$ of the Lambert W function in the interest of precision.
  \item The hybrid method proposed by Batzelis et al.~\cite{batzelisComputationLambertFunction2020}. This is a non-iterative method that aims to give a quick yet reasonable result at the cost of precision.
\end{itemize}

\begin{table}
    \centering
    \caption{Performance analysis of algorithms using the I-approach on solar cell SDMs with parameter sets taken from \cref{tab:model-parameter-values}.}
    \label{tab:time-performance-analysis-model-1}
    \begin{tabular}{cccc}
        \hline
        \textbf{Method} & \textbf{Mean time ($\mu$s)} & \textbf{Median time ($\mu$s)} & \textbf{RMSE} \\
        \hline
        \multicolumn{4}{c}{\textit{Parameter Set 1}} \\
        \hline
        Lambert W & 1704 & 1696 & --- \\
        Hybrid & 813 & 805 & $2.69\times10^{-3}$ \\
        Toledo et al. & 992 & 978 & $1.28\times10^{-15}$  \\
        LogWright & 874 & 865 & $1.44\times10^{-14}$  \\
        \hline

        \multicolumn{4}{c}{\textit{Parameter Set 2}} \\
        \hline
        Lambert W & 1695 & 1678 & --- \\
        Hybrid & 810 & 805 & $1.11\times10^{-4}$ \\
        Toledo et al. & 1017 & 998 & $2.14\times10^{-17}$ \\
        LogWright & 934 & 923 & $1.23\times10^{-15}$ \\
        \hline

        \multicolumn{4}{c}{\textit{Parameter Set 3}} \\
        \hline
        Lambert W & 1798 & 1793 & --- \\
        Hybrid & 841 & 839 & $7.03\times10^{-4}$ \\
        Toledo et al. & 1122 & 1112 & $3.84\times10^{-16}$ \\
        LogWright & 955 & 950 & $1.05\times10^{-15}$ \\
        \hline
        \multicolumn{4}{c}{\textit{Parameter Set 4}} \\
        \hline
        Lambert W & 1768 & 1759 & --- \\
        Hybrid & 830 & 821 & $7.47\times10^{-5}$ \\
        Toledo et al. & 986 & 979 & $7.02\times10^{-18}$ \\
        LogWright & 870 & 866 & $1.38\times10^{-15}$ \\
        \hline
        \multicolumn{4}{c}{\textit{Parameter Set 5}} \\
        \hline
        Lambert W & 1698 & 1684 & --- \\
        Hybrid & 810 & 804 & $1.04\times10^{-4}$ \\
        Toledo et al. & 993 & 980 & $1.95\times10^{-17}$ \\
        LogWright & 905 & 900 & $1.01\times10^{-15}$ \\
        \hline
        \multicolumn{4}{c}{\textit{Parameter Set 6}} \\
        \hline
        Lambert W & 1837 & 1823 & --- \\
        Hybrid & 847 & 843 & $8.42\times10^{-6}$ \\
        Toledo et al. & 1150 & 1145 & $5.22\times10^{-16}$ \\
        LogWright & 1102 & 1096 & $6.31\times10^{-16}$ \\
        \hline
    \end{tabular}
\end{table}

\begin{table}
    \centering
    \caption{Performance analysis of algorithms using the V-approach on solar cell SDMs with parameters taken from \cref{tab:model-parameter-values}. Baseline Lambert W isn't included for parameter sets 3-6 because it encounters numerical overflow during calculation.}
    % \caption{Performance analysis of algorithms based on computation time for V-approach. Parameter sets 3-6 from \cref{tab:model-parameter-values} is used. Baseline Lambert W isn't included because it encounters numerical overflow during calculation.}
    \label{tab:time-performance-analysis-model-2}
    \begin{tabular}{cccc}
        \hline
        \textbf{Method} & \textbf{Mean time ($\mu$s)} & \textbf{Median time ($\mu$s)} & \textbf{RMSE} \\
        \hline
        \multicolumn{4}{c}{\textit{Parameter Set 1}} \\
        \hline
        Lambert W & 1722 & 1701 & --- \\
        Hybrid & 776 & 769 & $1.76\times10^{-2}$ \\
        Toledo et al. & 1072 & 1060 & $2.08\times10^{-13}$ \\
        LogWright & 1030 & 1025 & $4.81\times10^{-13}$ \\
        \hline
        \multicolumn{4}{c}{\textit{Parameter Set 2}} \\
        \hline
        Lambert W & 1708 & 1686 & --- \\
        Hybrid & 778 & 772 & $1.38\times10^{-3}$ \\
        Toledo et al. & 1075 & 1065 & $4.81\times10^{-14}$ \\
        LogWright & 1046 & 1036 & $6.95\times10^{-14}$ \\
        \hline
        \multicolumn{4}{c}{\textit{Parameter Set 3}} \\
        \hline
        Hybrid & 810 & 807 & --- \\
        Toledo et al. & 921 & 915 & --- \\
        LogWright & 883 & 870 & --- \\
        \hline
        \multicolumn{4}{c}{\textit{Parameter Set 4}} \\
        \hline
        Hybrid & 802 & 797 & --- \\
        Toledo et al. & 917 & 905 & --- \\
        LogWright & 877 & 873 & --- \\
        \hline
        \multicolumn{4}{c}{\textit{Parameter Set 5}} \\
        \hline
        Hybrid & 775 & 772 & --- \\
        Toledo et al. & 1079 & 1071 & --- \\
        LogWright & 1056 & 1051 & --- \\
        \hline
        \multicolumn{4}{c}{\textit{Parameter Set 6}} \\
        \hline
        Hybrid & 813 & 808 & --- \\
        Toledo et al. & 894 & 885 & --- \\
        LogWright & 868 & 862 & --- \\
        \hline
    \end{tabular}
\end{table}

To test and compare performance, we generate I-V characteristics using each of the above approaches for 20000 iterations and compare the mean time and median time taken to generate I-V characteristics throughout those iterations. We also calculate the root mean square error (RMSE) of the generated I-V characteristics by considering the output of the SciPy implementation as the baseline. The above metrics are calculated for various parameter values taken from \cite{toledo2022quick} to ensure that we include cases in which the default Lambert W implementation encounters numerical overflow. All the parameter values used are listed in \cref{tab:model-parameter-values}. The iterative algorithms are configured such that their output precision is at least $10^{-8}$. To ensure a fair comparison, all the above algorithms are implemented and benchmarked in Python (version 3.11.3) on a machine with AMD Ryzen 7 5800H (3.2 GHz) processor and 16GB DDR4 RAM running the Manjaro Linux operating system. Our implementation of these algorithms has been posted to the GitHub repository\cite{LankireddyLogwrightGithub2023}.\\

All the performance analysis of the algorithms in terms of mean time, median time and RMSE is summarized in \cref{tab:time-performance-analysis-model-1,tab:time-performance-analysis-model-2}. Table~\ref{tab:time-performance-analysis-model-1} contains the metrics obtained when the I-approach is used to calculate the I-V characteristics for SDMs with parameter sets 1-6 taken from \cref{tab:model-parameter-values}, whereas \cref{tab:time-performance-analysis-model-2} contains the metrics when the V-approach is used for the same. As the Lambert W method encounters numerical overflow when calculating the I-V characteristics for models with parameters sets 3-6 using the V-approach, we neither include Lambert W in the results nor calculate the RMSE for any of the algorithms for those parameter sets.
From the tables, one can observe that the Hybrid algorithm take the least amount of time to run, in both the I-approach and the V-approach, closely followed by the LogWright method. However, the LogWright method gives an exact solution to the characteristic equation whereas the Hybrid method gives an approximate solution with error of the order $10^{-3}$. We also observe that I-approach is faster than the V-approach when working with parameter sets 1, 2, and 5, whereas V-approach seems to be faster than the I-approach when working with parameter sets 3, 4 and 6. No algorithm encounters numerical overflow when using the I-approach on any of the parameter sets. However, there are exceptions to the above observation, specifically in the cases of Hybrid method on parameter set 2 and Logwright method on parameter set 4.

\begin{figure*}[!t]
  \centering
  \subfloat[]{\includegraphics[width=2.8in]{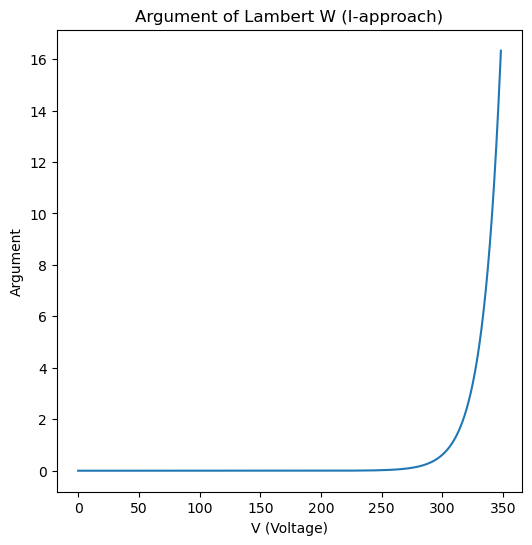}}
  \hfil
  \subfloat[]{\includegraphics[width=2.8in]{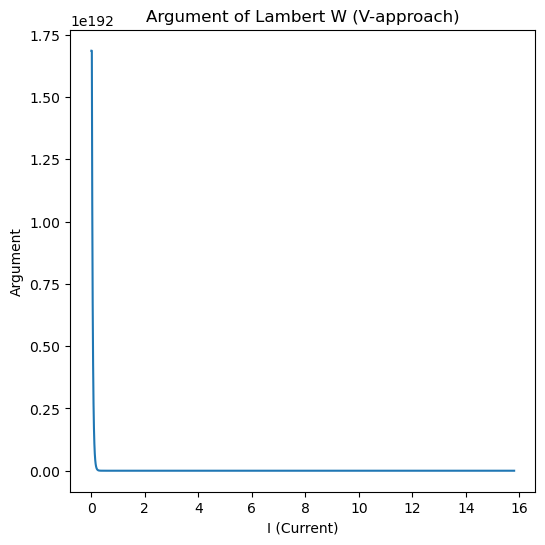}}
  \hfil
  \subfloat[]{\includegraphics[width=2.8in]{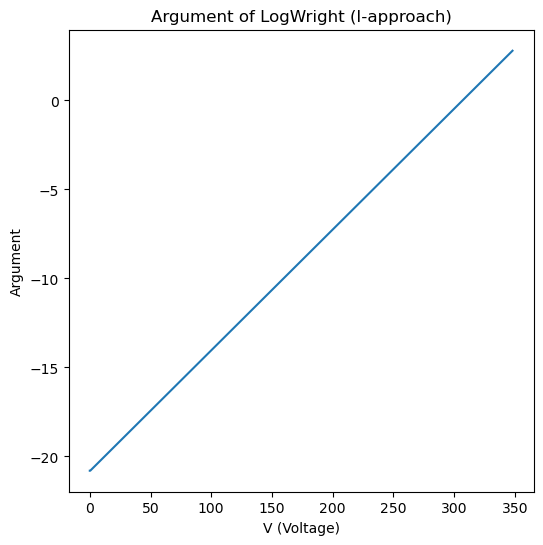}}
  \hfil
  \subfloat[]{\includegraphics[width=2.8in]{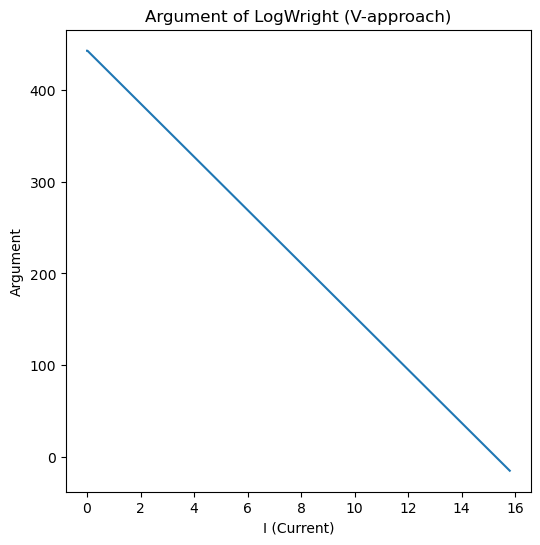}}
  \caption{Comparing arguments of the Lambert W function and the LogWright function when calculating I from V (Figures (a) and (c)) and when calculating I from V (Figures (b) and (d)). Figures (a) and (c) have been plotted from 0 to open circuit voltage, whereas Figures (b) and (d) have been plotted from 0 to the short circuit current for a solar cell, with SDM parameters as $I_{ph}=15.88A$, $I_{sat}=7.44 \times 10^{-10}A$, $a=14.67$, $R_s=2.04\Omega$ and $R_{sh}=425.2\Omega$.}
  \label{fig:comparing-function-arguments}
\end{figure*}

The values of arguments of the Lambert W function and the LogWright function that are obtained when calculating the I-V characteristics of a SDM (with parameters from set 1 of \cref{tab:model-parameter-values}) using the I-approach and V-approach are plotted in \cref{fig:comparing-function-arguments}. One can observe that the argument of Lambert W reaches magnitudes of the order of $10^{192}$ when using the V-approach on the parameter set 1 from \cref{tab:model-parameter-values}, whereas its counterpart in the LogWright function takes values that are a little over 400. This plot shows us how the LogWright function maps the arguments of the Lambert W function to the logarithmic scale. Doing so not only helps prevent numerical overflow, but also makes the computation faster.

\section{Conclusion}
\label{sec:conclusion}
\noindent The present work implements an effective method to obtain the current-voltage (I-V) characteristics of a solar cell represented using the single diode model and the LogWright function. This implementation based on the LogWright function, a logarithmic transformation of the Lambert W, solves the numerical overflow problems that one could encounter when using the Lambert W function to solve for the I-V characteristics. It also possesses properties which makes it desirable for field applications such as active load balancing. This implementation is compared with other methods that achieve the same purpose and is shown to be the fastest among the methods that give an exact solution, making it very useful for deployment in field installations.\\

There is a lot of potential for future work in exploring the relationship between solar cells and the LogWright function. Using the LogWright function can prevent numerical overflow issues wherever the argument given to the Lambert W function is of the form $x=a e^{b}$. One can study the usage of LogWright function in two-diode models and three-diode models, and also potential applications in modeling organic solar cells by studying \cite{roberts2015calculating} as a starting point. Tasks such as parameter estimation of diode models and maximum power point tracking can benefit from using the LogWright function.\\

Note: A preliminary version of this research was presented at the Canadian Association of Physics June-2022 annual conference, under the title "Solar Cells and the Lambert W Function", presented by S. R. Valluri on behalf of all co-authors.\\

% Part of this work was sponsored by a research grant from the Shastri Indo-Canadian Institute to the Indian Institute of Technology Tirupati and the University of Western Ontario.
Part of this work was sponsored by a research grant from the Shastri Indo-Canadian Institute to the Indian Institute of Technology Tirupati.
We would like to thank Dr. Prachi Kaul for her strong support in facilitating the Indo-Canadian research collaboration on this project.
S.R. Valluri is indebted to King's University College (UWO) for its consistent support of his research endeavors.

\bibliographystyle{IEEEtran}
\bibliography{refs}

% Biography text here without a photo.

% \begin{IEEEbiography}[{\includegraphics[width=1in,height=1.25in,clip,keepaspectratio]{fig1.png}}]{IEEE Publications Technology Team}
% In this paragraph you can place your educational, professional background and research and other interests.\end{IEEEbiography}

\end{document}